\documentclass[notitlepage,twocolumn,showpacs]{revtex4-1}
\usepackage[utf8]{inputenc}
\usepackage{amsmath}
\usepackage{amsfonts}
\usepackage{amssymb}
\usepackage{times,fullpage}
\usepackage{comment,color}
\usepackage{array,graphicx}

\graphicspath{{./figures/}}
 
\begin{document}

\newcommand{\nwc}{\newcommand}
\nwc{\vs}{\vspace}
\nwc{\hs}{\hspace}
\nwc{\la}{\langle}
\nwc{\ra}{\rangle}
\nwc{\nn}{\nonumber}
\nwc{\Ra}{\Rightarrow}
\nwc{\wt}{\widetilde}
\nwc{\lw}{\linewidth}
\nwc{\ft}{\frametitle}
\nwc{\ben}{\begin{enumerate}}
\nwc{\een}{\end{enumerate}}
\nwc{\bit}{\begin{itemize}}
\nwc{\eit}{\end{itemize}}
\nwc{\dg}{\dagger}
\nwc{\mA}{\mathcal A}
\nwc{\mD}{\mathcal D}
\nwc{\mB}{\mathcal B}
\nwc{\erfc}{\mbox{erfc}}

\nwc{\Tr}[1]{\underset{#1}{\mbox{Tr}}~}
\nwc{\pd}[2]{\frac{\partial #1}{\partial #2}}
\nwc{\ppd}[2]{\frac{\partial^2 #1}{\partial #2^2}}
\nwc{\fd}[2]{\frac{\delta #1}{\delta #2}}
\nwc{\pr}[2]{K(i_{#1},\alpha_{#1}|i_{#2},\alpha_{#2})}
\nwc{\av}[1]{\left< #1\right>}

\nwc{\zprl}[3]{Phys. Rev. Lett. ~{\bf #1},~#2~(#3)}
\nwc{\zpre}[3]{Phys. Rev. E ~{\bf #1},~#2~(#3)}
\nwc{\zpra}[3]{Phys. Rev. A ~{\bf #1},~#2~(#3)}
\nwc{\zjsm}[3]{J. Stat. Mech. ~{\bf #1},~#2~(#3)}
\nwc{\zepjb}[3]{Eur. Phys. J. B ~{\bf #1},~#2~(#3)}
\nwc{\zrmp}[3]{Rev. Mod. Phys. ~{\bf #1},~#2~(#3)}
\nwc{\zepl}[3]{Europhys. Lett. ~{\bf #1},~#2~(#3)}
\nwc{\zjsp}[3]{J. Stat. Phys. ~{\bf #1},~#2~(#3)}
\nwc{\zptps}[3]{Prog. Theor. Phys. Suppl. ~{\bf #1},~#2~(#3)}
\nwc{\zpt}[3]{Physics Today ~{\bf #1},~#2~(#3)}
\nwc{\zap}[3]{Adv. Phys. ~{\bf #1},~#2~(#3)}
\nwc{\zjpcm}[3]{J. Phys. Condens. Matter ~{\bf #1},~#2~(#3)}
\nwc{\zjpa}[3]{J. Phys. A ~{\bf #1},~#2~(#3)}
\nwc{\zpjp}[3]{Pramana J. Phys. ~{\bf #1},~#2~(#3)}

\title{Generalized Second Law and optimal protocols for nonequilibrium systems}
\author{Sourabh Lahiri$^{1,2}$}
\email{sourabhlahiri@gmail.com}

\author{Arun M. Jayannavar$^{3,4}$}
\email{jayan@iopb.res.in}

\author{Anupam Kundu$^1$}
\email{anupam.kundu@icts.res.in}

\affiliation{
  $^1$International Centre for Theoretical Sciences, TIFR, Survey no. 151, Sivakote Village, Hesaraghatta Hobli, Bengaluru 560089, India\\
  $^2$Birla Institute of Technology, Mesra, Ranchi, Jharkhand 835215, India\\
   $^3$Institute of Physics, Sachivalaya Marg, Bhubaneswar 751005, India\\
   $^4$Homi Bhabha National Institute, Training School Complex, Anushakti Nagar, Mumbai 400085, India.
 }

\begin{abstract}
 A generalized version of the Maximum Work Theorem is valid when the system is initially not at thermal equilibrium. In this work, we initially study the fraction of trajectories that violate this generalized theorem for a two simple systems: a particle in a harmonic trap (i) whose centre is dragged with some protocol, and (ii) whose stiffness constant changes as a function of time. We also find the optimal protocol that minimizes the average change in total entropy. To our surprise, we find that optimization of protocol does not necessarily entail maximum violation fraction.
 \end{abstract}

 \maketitle

 \section{Introduction}

 The thermodynamic Second Law has remained one of the most intriguing laws of the last two centuries. It provides a directionality to a spontaneous process. For instance, it is this law that tells us why broken shards of a vase do not join spontaneously to form the original vase, or why all the gas molecules in a room are not observed to congregate in one corner of the room. The Second Law provides constraints on what the maximum efficiency of a heat engine can be.   
 Ever since the pioneering work of Carnot in 1824, several equivalent formulations of this law has appeared in the literature, based on the works of Rudolf Clausius, Lord Kelvin, Max Planck, Constantin Carath\'eodory and many others \cite{kondepudi,kes}. One of them is the Maximum Work Theorem (MWT), which states that the maximum work that can be extracted from a process cannot exceed the work extracted in a reversible process. In our convention in this article, we would be using positive sign for work if it is done \emph{on} the system, and negative if it is extracted.
 For an isothermal, reversible process, the work done equals the change in Helmholtz free energy  of the system \cite{lan}. The mathematical statement of MWT would thus be $W>\Delta F$, where $\Delta F$ is the difference between the final and the initial free energies of the system.
 For any process where the system's initial and final states are at equilibrium, the MWT is equivalent to the more commonly used  statement of the Second Law that states that the total entropy of the universe (system and the surroundings with which it interacts) cannot decrease with time. The MWT is easier to verify in an experiment, because the work is a directly measurable quantity. 

 However, in practice, it is rarely the case that the system begins and ends in thermal equilibrium with its surroundings. For instance, we can consider a cyclic process in a time-periodic steady state \cite{lah12a_jpa}, and want to know what the Second Law means in terms of work done in a single cycle. Since the cyclic force drives the system out of equilibrium, the initial and final states of the system will be nonequilibrium ones. Another example would be that of memory erasure \cite{lut09_prl}, where the final state of a symmetric two-state memory device is in local equilibrium in one of the states, which is different from the global equilibrium in which both states are equally probable. Now one can ask the question: is the MWT still valid for such a process? Stated differently, is the work extracted in a reversible process still the maximum extractable work?
 This question has been addressed in a recent work \cite{esp11_epl}, where the authors have shown that the theorem must undergo a modification. In fact, instead of providing a lower bound to $W$, we obtain a lower bound on the total entropy change $\Delta S_{tot}$ of the system and its environment (see Eq. \eqref{MET}): $\Delta S_{tot}\ge 0$. Let us term this theorem as minimum entropy theorem (MET). It is this theorem that is equivalent to the statement of non-negativity of the change in total entropy.

 We will begin by examining the MWT for a small system in contact with a thermal reservoir, and whose dynamics is dominated by thermal noise. Such systems have invited a lot of attention in recent years, because the improvement of experimental techniques to handle small systems like individual macromolecules or colloidal particles have led access to very precise measurement of work done on them \cite{rit05_pt}.
 On the theoretical side, a bunch of relations collectively known as the ``Fluctuation Theorems'' (FTs) have come into existence \cite{jar11_jcm,sei12_rpp,han11_rmp}. These provide stringent symmetry conditions on the probability distributions of thermodynamic variables like work or entropy. The FTs involving work are also referred to as nonequilibrum work theorems in the literature \cite{sei12_rpp}. One of these theorems, called the Jarzynski Equality \cite{jar97_prl,jar97_pre}, states the following. If the systems begins in a state of thermal equilibrium with its environment and is thereafter perturbed by a time-dependent protocol, then the work done $W$ during the process is related to the change in free energy through the following exact relation:
 \begin{align}
   \av{e^{-\beta W}} = e^{-\beta\Delta F}.
 \end{align}
  The angular brackets denote averaging over an ensemble of experimental realizations of the process. The equality holds even when the system has been driven far from equilibrium, and so is stronger than the linear response theory which in turn is valid only for small perturbations about equilibrium. A direct corollary of the Jarzynski Equality is the MWT:
\begin{align}
  \av{W}\ge \Delta F,
  \label{MWT}
\end{align}
which is obtained by application of the Jensen's inequality. We note that the averaging is essential for a small system, since the thermodynamic variables are rendered stochastic by the thermal noise. Both work $W$ and total entropy change $\Delta s_{tot}$ will in general vary from one experimental realization to another. Thus, the MET will now be given by
\begin{align}
  \av{\Delta s_{tot}}\ge 0.
\end{align}
We will find later that even for this case (initial state at equilibrium), the MWT gets modified to MET if the final state is out of equilibrium, although this effect is not captured by the Jarzynski Equality. Relation \eqref{MWT} and the MET are equivalent if, and only if, \emph{both} the initial and the final states are at thermal equilibrium.
Let us first consider this special case.
Since the inequality involves averaged work, work done along individual trajectories can ``violate'' this inequality. In other words, the work distribution can have a portion in which $W<\Delta F$, although its mean must follow \eqref{MWT}. If we can find a protocol  that minimizes the average dissipated work $\av{W}-\Delta F$ with given initial and final values of the parameter, then this protocol would be more efficient than the others.
Of course, if there is no constraint on the time of observation, we know that the most efficient protocol is the one where $\tau\to\infty$ \emph{and} the variation of the parameter is slow enough so as to make the process quasistatic. This is because, as stated before, the maximum amount of work can be extracted from the system in such a process. However, if we have a fixed time of observation, the form of the optimal protocol is non-trivial, as has been shown in \cite{sei07_prl}. In particular, the authors of \cite{sei07_prl} showed that the optimal protocol consists of finite jumps at the beginning and at the end of the process, at least for the harmonic confining potential that they have considered.

Now, as mentioned before, the MWT and MET are inequivalent if the system is initially and/or finally not at thermal equilibrium. In that case, as shown in \cite{esp11_epl},  we can have mean work less than the change in equilibrium free energy. The exact relation is given by (setting the Boltzmann constant to be $k_B=1$ hereafter)
\begin{align}
  \av{W}-\Delta F = T\Delta_iS + T\Delta I.
  \label{gen_sec_law}
\end{align}
Here, $\Delta_iS$ is the net irreversible contribution to the system's entropy change, which is always non-negative. The net entropy change of the system consists of, in addition to $\Delta_iS$, the entropy change $\Delta_eS$ due to the reversible transfer of energy between the system and its environment, so that $\Delta S = \Delta_iS + \Delta_eS$. The second quantity on the right hand side is  $\Delta I = I(\tau)-I(0)$, where $I(t)$ is the Kullback-Leibler divergence between the actual distribution reached at time $t$ and the corresponding equilibrium distribution:
\begin{align}
 I(t) = D[p(t)||p^{eq}(t)] = \int dx ~p(x,t)\ln\frac{p(x,t)}{p^{eq}(x,t)}.
\end{align}
Now, $\Delta_iS$ being non-negative, Eq. \eqref{gen_sec_law} gives rise to the inequality
\begin{align}
  \av{W}-\Delta F \ge T\Delta I.
  \label{gen_sec_law1}
\end{align}
Defining $A\equiv \av{W}-T\Delta I$, we obtain the MET:
\begin{align}
  A\ge \Delta F.
  \label{MET}
\end{align}
Eq. \eqref{MET} is equivalent to MET, as we will show below. However, it is not the same as MWT, as we can observe by comparing equations \eqref{MWT} and \eqref{MET}.

Let us now define the stochastic versions of the quantities arising in the right hand side of Eq. \eqref{gen_sec_law}, that are calculable for each stochastic trajectory $X$.  These stochastic variables $\Delta_is[X]$ and $i(x,t)$ are defined such that $\av{\Delta_i s[X]} = \Delta_i S$ and $\av{i(x,t)} = I(t)$, where $\Delta_i s[X]$  is a functional of the full trajectory $X$.
With these definitions, we can rewrite the detailed form of \eqref{gen_sec_law} that is applicable to each trajectory:
\begin{align}
  W - \Delta F = T\Delta_is + T\Delta i.
\end{align}
Of course, on averaging both sides of the above equation, we get back Eq. \eqref{gen_sec_law}. 

We note from these definitions that $i(x,t) \equiv \ln[p(x,t)/p^{eq}(x)]$, and that for an ideal heat bath $\Delta_is$ is simply the change in the stochastic total entropy $\Delta s_{tot}$ in a realization \cite{sei05_prl,sei08_epjb,sei12_rpp}. In our models, we will always assume that this is the case. This total change in entropy follows an integral Fluctuation Theorem, as was shown in \cite{sei05_prl}:
\begin{align}
  \av{e^{-\Delta s_{tot}}} = 1,
\end{align}
which, on application of Jensen's inequality, trivially gives $\Delta_iS = \av{\Delta s_{tot}}\ge 0$, as mentioned before.

Our objective in this paper is to study this inequality in the light of the violation fraction, i.e. the fraction of trajectories in which we obtain $W-\Delta F - T\Delta i < 0$. This is equivalent to computing the fraction of trajectories having $\Delta s_{tot} < 0$. We will further find an optimal protocol that minimizes $\av{\Delta s_{tot}}$, and compare it with the optimal protocol that minimizes only the average work done \cite{sei07_prl}.
We find that the optimal protocol so obtained is different from the one obtained in \cite{sei07_prl} \emph{even when the initial state is at equilibrium}. When only the shape of the potential is changed by the protocol (e.g. changing stiffness constant of a harmonic trap), The optimal protocol is found to be simply the one in which the external parameter remains constant throughout the process from time $t=0^+$ to $t=\tau^-$, with a jump at the final time only (if the system is initially at equilibrium). If the system is initially in a non-equilibrium state, then the optimal protocol involves jumps both at the initial and the final times (the parameter remaining frozen in-between).
We, surprisingly, find that although the average of the change in total stochastic entropy is higher in \cite{sei07_prl} than in the case of the optimal protocol obtained by us, the violation fraction (to be defined later) for this entropy can be higher in the former case, which is quite counter-intuitive.

\section{Derivation of the generalized Second Law}

According to stochastic thermodynamics \cite{sek98,sei12_rpp}, one can have definitions of thermodynamic quantities like work, heat or entropy for a single realization of a process. Let us consider the Langevin equation for an overdamped particle of mass $m$ that is moving in a medium of inverse temperature $\beta$ and friction coefficient $\gamma$:
\begin{align}
  \gamma \dot x =  - \partial_x V(x,t) + \xi(t).
\end{align}
Here, the overhead dot implies derivative with respect to time. $\xi(t)$ is the thermal noise in the medium, which will be assumed to be delta-correlated ($\av{\xi(t)\xi(s)} = 2(\gamma/\beta)\delta(t-s)$) and following a Gaussian distribution of vanishing mean. $V(x,t)$ is the net potential acting on the particle, and hence $-\partial_xV$ is the net force the particle has been subjected to, other than the forces generated by the heat bath. We will not consider the effect of non-conservative forces in this article, although the generalization to such cases is straightforward \cite{sei12_rpp}. 

After some simple algebraic manipulations of the Langevin equation, one can identify \cite{sek98}:
\begin{align}
  E(\tau) &\equiv V(x_\tau,\tau) &&\mbox{(Internal energy)}\nn\\
  W(\tau) &\equiv \int_0^\tau \pd{E(x,t)}{t}dt &&\mbox{(Work done)}\nn\\
  Q(\tau) &\equiv \int_0^\tau (\gamma\dot x - \xi(t))\circ \dot x dt &&\mbox{(Dissipated heat)}.
\end{align}
The $\circ$ symbol in the definition of heat implies that the product is of Stratonovich type \cite{gar}. 

The relations appearing in the previous section can be easily derived by noting that a \emph{nonequilibrium} free energy can be defined as $F^{neq}(x,t) = E(x,t) - Ts(x,t)$, where $E(x,t)$ and $s(x,t)$ are  the energy and the entropy of the system. As shown by Seifert \cite{sei05_prl,sei08_epjb,sei12_rpp}, the definition of such nonequilibrium entropy for a single realization at time $t$ is given by
\begin{align}
  s(x,t) \equiv -\ln p(x,t).
  \label{s}
\end{align}
where $p(x,t)$ is the distribution of the particle position at time $t$, which is a solution of the Smoluchowski equation.

Using the First Law, $W = Q + \Delta E$, we find for a given trajectory $X$:
\begin{align}
  W[X]-\Delta F^{neq} &= Q[X] + T\Delta s = Q[X] + T(\Delta_is + \Delta_es) = T\Delta_i s,
  \label{stot}
\end{align}
using the fact that $\Delta_e s = -Q/T$.
For an ideal heat bath, $\Delta_is = \Delta s_{tot}$. Now, the equilibrium distribution corresponding to the external parameters at time $t$ is given by $p^{eq}(x,t) = e^{-\beta(E(x,t)-F)}$, where $F$ is the equilibrium free energy. It can be rewritten as $E(x,t) = F - T\ln p^{eq}(x,t)$. Then we get (using Eq. \eqref{s})
\begin{align}
  F^{neq}(x,t) = F + T\ln\frac{p(x,t)}{p^{eq}(x,t)} = F + T i(x,t).
\end{align}
Thus, \eqref{stot} reduces to
\begin{align}
  W - \Delta F^{neq} = W - \Delta F - T \Delta i = T\Delta s_{tot}.
  \label{trajectory_equation}
\end{align}
Further, using the fact that $\Delta S_{tot} \equiv \av{\Delta s_{tot}}\ge 0$, we obtain the inequality \eqref{gen_sec_law1}.

\section{Violation fraction for a dragged overdamped particle in a harmonic trap}

We note that the quantity $\av{W}-\Delta F - T\Delta I$ is always non-negative. However, if the averaging is not done, then $W-\Delta F - T\Delta i$ can be negative for individual realizations.  We call them ``atypical realizations''. The violation fraction is the fraction of such atypical realizations occuring in an ensemble of realizations of a process. In other words, it is the probability of observing an event that is not observed in macroscopic systems. The violation fraction for work for a particle in a harmonic trap as well as in a double-well potential has been studied in \cite{lah11_jpa}, when the initial distribution is a thermal one. A related quantity has been studied in \cite{gar14_pre,gar14a_pre}. If the violation fraction is high, we may naively expect the process to be more efficient. As will be discussed in sec. \ref{subsec:k_t}, this is not always true. A process can have both higher violation fraction as well as higher dissipation.

\subsection{Violation fraction for arbitrary dragging protocol}

Let us consider a colloidal particle in a harmonic trap, initially at a nonequilibrium (but Gaussian) distribution with zero mean. At time $t=0$ we switch on the protocol, which consists of dragging the centre of the trap according to a protocol $\lambda(t)$. The overdamped Langevin equation is given by
\begin{align}
  \gamma\dot x = -k(x-\lambda(t))+\xi(t),
  \label{Lang_drag_arbitrary}
\end{align}
where $k$ is the spring constant of the harmonic trap, $\gamma$ is the friction coefficient, and $\xi(t)$ is the thermal noise which follows a Gaussian distribution with mean zero, and which is delta-correlated in time: $\av{\xi(t)\xi(s)} = 2\gamma T\delta(t-s)$, where $T$ is the temperature of the heat bath.

From \eqref{Lang_drag_arbitrary}, we have
\begin{align}
  x(t) = x_0e^{-kt/\gamma} + \frac{1}{\gamma}\int_0^t ds [k\lambda(s)+\xi(s)]e^{-k(t-s)/\gamma}.
\end{align}
Since $x$ is linear in the initial position and the noise, both of which are Gaussian, it must follow a Gaussian distribution itself.

The work done is given by
\begin{align}
  W = -k\int_0^\tau dt (x-\lambda(t))\dot\lambda(t).
\end{align}
Being linear in $x$, the work is also a Gaussian variable. We therefore need to compute only the mean and the variance of $W$. The mean position at time $t$ is given by:
\begin{align}
  \av{x(t)} &= \frac{k}{\gamma}\int_0^t ds\lambda(s)e^{-k(t-s)/\gamma}.
              \label{xmean}
\end{align}
The mean and variance of the work are then
\begin{align}
  \av{W} &= -k\int_0^\tau dt [\av{x(t)}-\lambda(t)]\dot\lambda(t)
           \label{Wmean}
\end{align}
and
\begin{align}
  \sigma_W^2 &= k^2\int_0^\tau dt\int_0^\tau ds \{\av{[x(t)-\av{x(t)}][x(s)-\av{x(s)}]}\nn\\
  &\hspace{5cm}\times\dot\lambda(t)\dot\lambda(s)\}.
               \label{Wvar}
\end{align}
Using the given form of $\lambda(t)$ in \eqref{xmean}, and substituting this equation in \eqref{Wmean} and \eqref{Wvar}, we can arrive at the distribution for $W$.

Now, the violation of generalized second law by work done can now be readily computed. The fraction of trajectories along which $W$ is less than $\Delta F+T\Delta I$ will constitute the violation fraction $f_W$. From the definition of the violation fraction ($f_W$) as mentioned at the beginning of this subsection, we have
\begin{align}
  f_W = \int_{-\infty}^{\Delta F+T\Delta I}P(W)dW.
\end{align}
Using the Gaussian form of $P(W)$ and the fact that $\Delta F=0$ for a dragged particle, we then have
\begin{align}
  f_W &= \frac{1}{2\pi\sigma_W^2}\int_{-\infty}^{T\Delta I}\exp\bigg[-\frac{(W-\av{W})^2}{2\sigma_W^2}\bigg]dW \nn\\
      &= \frac{1}{2}\erfc\bigg[\frac{\av{W}-T\Delta I}{\sqrt{2\sigma_W^2}}\bigg].
        \label{ana_viol}
\end{align}

\subsection{Special case: centre of trap is dragged with uniform velocity}

Let us consider a situation in which the centre of the trap is dragged at uniform velocity $v_1$. Our protocol $\lambda(t)$ in the previous section then becomes $\lambda(t) = v_1t$.  The equation of motion followed by the particle is assumed to be given by the overdamped Langevin equation:
\begin{align}
  \gamma\dot x = -k(x-v_1t)+\xi(t),
  \label{Lang_drag}
\end{align}
 The formal solution of Eq. \eqref{Lang_drag} is given by
\begin{align}
  x(t) = x_0 e^{-kt/\gamma}+\frac{1}{\gamma}\int_0^t ds[kv_1s+\xi(s)]e^{-k(t-s)/\gamma}.
  \label{Lang_sol}
\end{align}
Here, $x_0$ is sampled from the initial Gaussian distribution with mean zero and variance $\sigma_0^2$.
Since $x(t)$ is linear in $x_0$ and the noise, which are Gaussian variables, it itself must follow a Gaussian distribution.

%
We can readily obtain from Eq. \eqref{Lang_sol}
\begin{align}
  \av{x(t)} &= \frac{kv_1}{\gamma}e^{-kt/\gamma}\int_0^t ds~se^{ks/\gamma} \nn\\
  &= v_1t - \frac{\gamma v_1}{k}[1-e^{-kt/\gamma}].
\end{align}
Using this relation  we get
\begin{align}
  \av{W} = \gamma v_1^2\bigg[\tau-\frac{\gamma}{k}\big(1-e^{-k\tau/\gamma}\big)\bigg].
  \label{Wavg2}
\end{align}
The variance of $W$ is given by
\begin{align}
  \sigma_W^2 &= \gamma^2v_1^2\bigg(\sigma_0^2-\frac{T}{k}\bigg)\big(1-e^{-k\tau/\gamma}\big)^2 \nn\\
  &\hspace{1cm}+ 2v_1^2\gamma T\bigg[\tau-\frac{T}{k}\big(1-e^{-k\tau/\gamma}\big)\bigg].
               \label{Wvar2}
\end{align}
Thus, Eqs. \eqref{Wavg2} and \eqref{Wvar2} specify the full distribution for work.

\begin{figure}[!h]
  \centering
  \includegraphics[width=8cm]{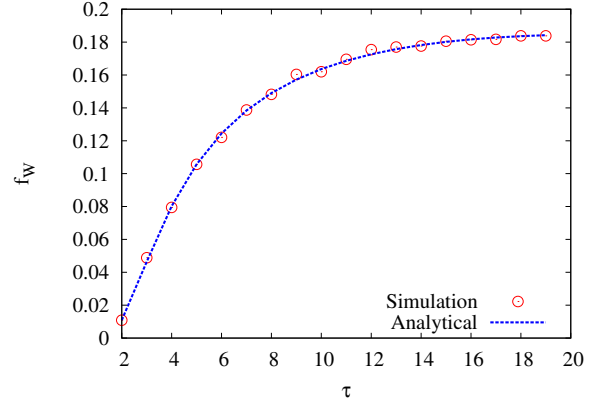}
  \caption{Plot of violation fraction of work (i.e. fraction of trajectories in which $W<T\Delta I$) as a function of the time of observation. The blue curve is the analytical result given by \eqref{ana_viol}, and the red open circles are the ones generated by simulation of the system. The parameters used are: $k=1$, $\sigma_0^2=0.1$, $T=0.5$, $v_1=0.1$ and $\gamma=1$.}
  \label{fig:ana_viol_tau}
\end{figure}

The violation fraction is given by (see eq. \eqref{ana_viol})
\begin{align}
  f_W =\frac{1}{2}\erfc\bigg[\frac{\av{W}-T\Delta I}{\sqrt{2\sigma_W^2}}\bigg].
  \label{ana_viol1}
\end{align}
The initial and final values of the information are given by
\begin{align}
  I(0) &= \frac{1}{2}\ln \frac{T}{k\sigma_0^2}+\frac{k\sigma_0^2}{2T}-\frac{1}{2}; \nn\\
  I(\tau) &= \frac{1}{2}\ln \frac{T}{k\sigma_0^2}+\frac{k}{2T}\big[\av{x_\tau}^2-2v_1\tau\av{x_\tau}\nn\\
  &\hspace{2cm}+\sigma_\tau^2+v_1^2\tau^2\big]-\frac{1}{2}.
            \label{Delta_I}
\end{align}
The variance in position is given by
\begin{align}
  \sigma_\tau^2 = \bigg(\sigma_0^2-\frac{T}{k}\bigg)e^{-2k\tau/\gamma}+\frac{T}{k}.
  \label{xvar}
\end{align}
Using \eqref{Wavg2}, \eqref{Wvar2}, \eqref{Delta_I} and \eqref{xvar}  in \eqref{ana_viol}, we can compute the violation fraction $f_W$.
Note that since the argument of the complementary error function is always positive, the violation fraction cannot exceed 1/2. Eq. \eqref{ana_viol1} has numerically been tested in figure \ref{fig:ana_viol_tau}, where we have plotted $f_W$ as a function of the observation time $\tau$. We find that the simulated curve agrees with the analytical one to a very good accuracy. We note from the figure that in contrast to the case where the initial distribution is an equilibrium one \cite{lah11_jpa}, the violation fraction \emph{increases} with increase in the time of observation.

\section{Optimal protocol for particle in a harmonic trap}

We now proceed to find which protocol minimizes the quantity $\av{W}-\Delta F -T\Delta I$ for a given time of observation $\tau$, and for given initial and final values of the parameter $\lambda_0$ and $\lambda_\tau$, respectively. We will follow the approach of Schmeidl and Seifert \cite{sei07_prl}, where the form of protocol that minimizes the mean work (only) was derived. We will compare the optimal protocols obtained below with the one obtained in \cite{sei07_prl}, and show that the results are quite different even when the system is initially at thermal equilibrium with the bath.

\subsection{Dragging the centre of the trap}

Once again, the particle is placed in a harmonic potential whose centre is moved according to some protocol $\lambda(t)$. In other words, the particle is placed in a potential $V(x,t) = \frac{1}{2}k[x-\lambda(t)]^2$. We assume $k=T=\gamma=1$, so that the Langevin equation becomes
\begin{align}
  \dot x = \lambda(t)-x+\xi(t),
\end{align}
We note that the average position depends on the form of the protocol, but not the variance: $\sigma^2(t) = 1+(\sigma_0^2-1)e^{-2t}$, as can be shown from the Langevin equation. This prompts us to use the variable $u\equiv\av{x}$ to rewrite the Langevin equation as  $\dot u(t)=\lambda(t)-u(t)$, where $u\equiv\av{x}$.
The mean work is obtained to be \cite{sei07_prl}
\begin{align}
  \av{W} = \int_0^\tau dt ~{\dot u}^2(t) + \frac{1}{2}({\dot u}^2_\tau-{\dot u}^2_0).
\end{align}
$\Delta I$ is given by
\begin{align}
  \Delta I &= \frac{1}{2\sigma_\tau}(\dot u_\tau^2+\sigma_\tau^2-1)-\frac{1}{2\sigma_0}(\sigma_0^2-1).
\end{align}
Here, we have used the relation $\lambda_\tau-u_\tau = \dot u_\tau$.
We define the variable
\begin{align}
  A &\equiv \av{W}-\Delta I\nn\\
    & = \int_0^\tau dt ~{\dot u}^2(t) + \frac{1}{2}({\dot u}^2_\tau-{\dot u}^2_0)\nn\\
  &\hspace{1cm}-\frac{1}{2\sigma_\tau}(\dot u_\tau^2+\sigma_\tau^2-1)+\frac{1}{2\sigma_0}(\sigma_0^2-1).
\end{align}
Noting that other than the first term, all others involve boundary terms, the Euler-Lagrange equation will be the same as in \cite{sei07_prl}: $\ddot u=0$. With $u_0=0$, the solution is
\begin{align}
  u(t)=mt,
\end{align}
where $m$ is a constant. The other boundary condition is (choosing $\lambda_0=0)$, $\dot u_0 = \lambda_0-u_0=0$. With this form of $u(t)$, the expression for $A$ becomes
\begin{align}
  A &= m^2\tau + \frac{1}{2}(\lambda_\tau-m\tau)^2 \nn\\
  &- \frac{1}{2\sigma_\tau}[(\lambda_\tau-m\tau)^2+\sigma_\tau^2-1] + \frac{1}{2\sigma_0}(\sigma_0^2-1).
\end{align}
We want to minimize $A$ with respect to $m$. This gives
\begin{align}
  &\pd{A}{m}\bigg|_{m=m^*} = 0\nn\\
  \Ra~ & m^* = \frac{\lambda_\tau\big(1-\frac{1}{\sigma_\tau}\big)}{2+\tau\big(1-\frac{1}{\sigma_\tau}\big)}.
\end{align}
With this choice, the optimal value of $A$ becomes
\begin{align}
  A^* = \frac{\lambda_\tau^2(\sigma_\tau-1)}{2\sigma_\tau + \tau(\sigma_\tau-1)} + \frac{1}{2}\bigg(\sigma_0-\frac{1}{\sigma_0}-\sigma_\tau+\frac{1}{\sigma_\tau}\bigg),
\end{align}
where
\begin{align}
  \sigma_\tau^2 = 1+(\sigma_0^2-1)e^{-2\tau},
\end{align}
which is obtained from the Langevin equation.

The optimal protocol becomes
\begin{align}
  \lambda^*(t) &= \dot u^*(t) + u^*(t)= m^*(1+t) \nn\\
               &= \frac{\lambda_\tau\big(1-\frac{1}{\sigma_\tau}\big)(1+t)}{2+\tau\big(1-\frac{1}{\sigma_\tau}\big)}.
\end{align}
We therefore observe that although the form of the protocol is linear as in \cite{sei07_prl}, the exact expression is different.

\subsection{Changing the stiffness constant of the trap}
\label{subsec:k_t}

Let the stiffness constant of the trap be varied according to the protocol $\lambda(t)$. In other words, the time-dependent potential is given by $V(x,t) = \frac{1}{2}\lambda(t)x^2$. The corresponding Langevin equation is
\begin{align}
  \dot x = -\lambda(t)x+\xi(t).
\end{align}
This time, in contrast to the previous case, the mean position does not depend on the functional form of $\lambda(t)$, but the variance does. Thus we define the new variable $w\equiv \av{x^2}$, so that the Langevin equation becomes
\begin{align}
  \dot w = -2\lambda w + 2.
  \label{lang_w}
\end{align}
Again, as shown in \cite{sei07_prl}, the mean work can be written in terms of the variable $w$:
\begin{align}
  \av{W} &= \frac{1}{4}\int_0^\tau dt \frac{\dot w^2}{w} + \frac{1}{2}(\lambda_\tau w_\tau-\lambda_0 w_0) - \frac{1}{2}\ln\frac{w_\tau}{w_0}.
\end{align}
In order to calculate $\Delta I$, we note that the distributions are given by
\begin{align}
  p_i(x_0) &= \frac{e^{-x_0^2/2\sigma_0^2}}{\sqrt{2\pi\sigma_0^2}}; \hspace{1cm} p_i^{eq}(x_0)= \sqrt{\frac{\lambda_0}{2\pi}}~e^{-\lambda_0 x_0^2/2}; \nn\\
  p_f(x_\tau) &= \frac{e^{-x_\tau^2/2\sigma_\tau^2}}{\sqrt{2\pi\sigma_\tau^2}}; \hspace{1cm} p_f^{eq}(x_\tau)= \sqrt{\frac{\lambda_\tau}{2\pi}}~e^{-\lambda_\tau x_\tau^2/2}.
\end{align}
Therefore, since $\Delta F = \frac{1}{2}\ln\big(\frac{\lambda_\tau}{\lambda_0}\big)$ is independent of $w_0$ and $w_\tau$, we again consider the quantity
\begin{align}
  A \equiv \av{W}-\Delta I &= \frac{1}{4}\int_0^\tau dt \frac{\dot w^2}{w} + \frac{1}{2}(\lambda_\tau w_\tau-\lambda_0 w_0) \nn\\
                           &- \frac{1}{2}\ln\frac{w_\tau}{w_0}- \frac{1}{2}[\lambda_\tau w_\tau - \ln(\lambda_\tau w_\tau)-1]\nn\\
  &+ \frac{1}{2}[\lambda_0 \sigma_0^2 - \ln(\lambda_0 \sigma_0^2)-1].
\end{align}
Once again, the Euler-Lagrange equations are the same as in \cite{sei07_prl}, and are given by
\begin{align}
  \dot w^2 - 2w\dot w = 0,
\end{align}
whose general solution is
\begin{align}
  w(t) = c_1(1+c_2t)^2.
\end{align}
The initial condition gives $w_0=\sigma_0^2~\Ra~c_1=\sigma_0^2$. Thus, we have
\begin{align}
  w = \sigma_0^2(1+c_2t)^2; \hspace{1cm} \dot w = 2c_2\sigma_0^2(1+c_2t).
\end{align}
Then the expression for $A$ becomes
\begin{align}
  A &= \frac{\sigma_0^2}{\tau}(b-1)^2 + \frac{1}{2}[\lambda_\tau \sigma_0^2 b^2-\lambda_0\sigma_0^2] - \ln b\nn\\
    &\hspace{2cm} -\frac{1}{2}[\lambda_\tau \sigma_0^2 b^2 - \ln(\lambda_\tau \sigma_0^2 b^2) - 1] \nn\\
  &\hspace{2cm}+ \frac{1}{2}[\lambda_0 \sigma_0^2 - \ln(\lambda_0 \sigma_0^2)-1] \nn\\
  &= \frac{\sigma_0^2}{\tau}(b-1)^2 + \frac{1}{2}\ln\bigg(\frac{\lambda_\tau}{\lambda_0}\bigg),
\end{align}
where $b=1+c_2\tau$. Thus, the value of $b$ that minimizes $A$ is $b^* = 1$, which implies $c_2^* = 0$.
Correspondingly, $A^* = \frac{1}{2}\ln\big(\frac{\lambda_\tau}{\lambda_0}\big) = \Delta F$, or $\av{W} - \Delta I - \Delta F = \Delta S_{tot} = 0$ (see Eq. \eqref{gen_sec_law}).
We further have $w^*(t) = c_1 = \sigma_0^2$, which is independent of time. Finally, we get
\begin{align}
  \lambda^*(t) = \frac{2-\dot w}{2w}\bigg|_{c_2=c_2^*} = \frac{1}{\sigma_0^2},
  \label{optimal_protocol}
\end{align}
which is  a constant protocol with jumps at the two ends: at time $t=0^+$, the value of the external parameter is switched instantaneously from $\lambda(0) = \lambda_0$ to $\lambda(0^+) = 1/\sigma_0^2$, while at the final time, it is switched from $\lambda(\tau^-) = 1/\sigma_0^2$ to $\lambda(\tau) = \lambda_\tau$. This is understandable on the basis of the following physical argument: due to the intial quench, the nonequilibrium initial distribution becomes a thermal distribution with respect to the new potential. Thus, the system does not undergo any relaxation, so neither the distribution of states changes nor is any heat dissipated in the process. This means that both the system's entropy as well as the medium's entropy do not change, so that the total entropy remains constant. At the final instant, due to the sudden quench, once again the total entropy does not change. Thus, $\Delta S_{tot}=0$ in the entire process, although the process is not quasi-static. We can the infer that this can be generalized to potentials that are not harmonic, and apply a similar protocol to get vanishing change in total entropy.

Evidently, if the initial distribution is the thermal one: $\sigma_0^2 = 1/\lambda_0$, then this means that we do not perturb the system at all from time $t=0$ to $t= \tau^-$, and at the final instant perform a sudden quench by changing the value of the parameter to $\lambda_\tau$.

\begin{figure}[!h]
  \centering
  \includegraphics[width=8cm]{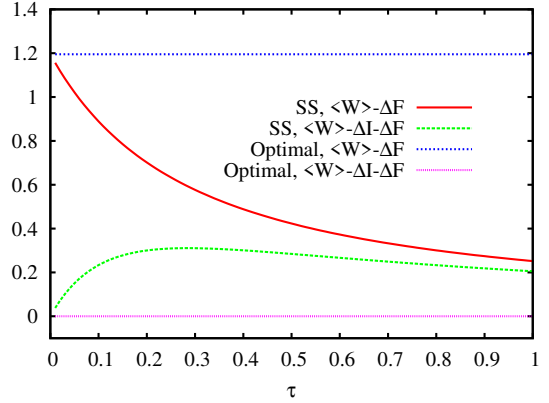}
  \caption{Variation of mean work and $A$ as functions of the observation time $\tau$, for SS protocol and the optimal protocol. The parameters used are: $\lambda_0 = 1$, $\lambda_\tau =5$.}
  \label{fig:optimal}
\end{figure}

Let us check how the results from Schmeidl and Seifert's protocol compare with the optimal protocol we obtained above. In figure \ref{fig:optimal}, we have plotted the variations of the dissipated work $\av W-\Delta F$ and $\Delta S_{tot} \equiv \av W - \Delta I - \Delta F$, with the time of observation $\tau$. Since the SS protocol is tailored to obtain minimum work, we find that the red curve is always below the blue curve. However, since the optimal protocol is tailored to obtain the minimum value of $\Delta S_{tot}$, the green curve (for Seifert's protocol) always lies above the magenta curve. Note that in the case of the optimal protocol, the mean work $\av{W}$ is given by the summation of the changes in internal energy at the beginning and at the end of the protocol (see Eq. \eqref{W_optimal} below), and is independent of the time of observation. So is $\Delta S_{tot}$, because it is always zero for the optimal protocol.

\begin{figure}[!h]
  \centering
  \includegraphics[width=8cm]{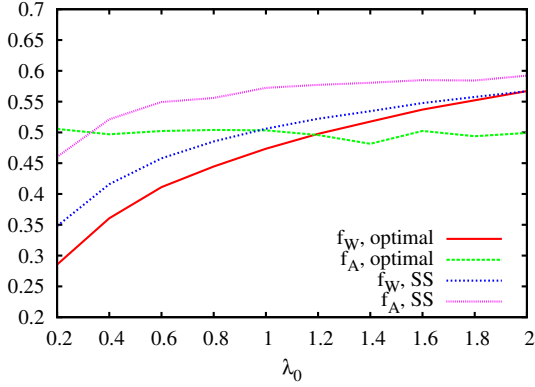}
  \caption{Variation of violation fractions with respect to $W$ (i.e. $W<\Delta F$) and $A$ (i.e. $W-\Delta i<\Delta F$, where $\Delta i= \ln\frac{p(x_\tau,\tau)}{p^{eq}(x_\tau)} - \ln\frac{p(x_0,0)}{p^{eq}(x_0)}$) as functions of initial value $\lambda_0$ of the stiffness constant. We have chosen $\lambda_\tau=5$ and $\tau=1$ in the simulations.}
  \label{fig:viol_rk0}
\end{figure}

In figure \ref{fig:viol_rk0}, the plots of the violation fractions for work and $W-\Delta I$ have been plotted as functions of the initial value $\lambda_0$ of the stiffness constant. We find that although the violation fractions $f_W$ with respect to work (i.e. the fraction of trajectories violating the inequality $W\ge \Delta F$) intersect each other at around $\lambda_0 = 1.8$, the violation fraction $f_A$ (i.e. the fraction of trajectories violating the inequality $W-\Delta I\ge \Delta F$)  can be less for the optimal protocol (for which it is pegged at 0.5) than the $f_A$ for the SS protocol.

\begin{figure}[!h]
  \centering
  \includegraphics[width=8cm]{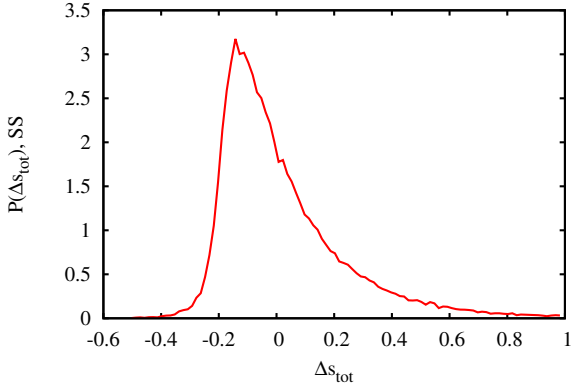}
  \caption{Distribution of $\Delta s_{tot} = W-\Delta F - \Delta i$ for the SS protocol, where $\Delta i= \ln\frac{p(x_\tau,\tau)}{p^{eq}(x_\tau)} - \ln\frac{p(x_0,0)}{p^{eq}(x_0)}$, where the value of initial stiffness constant has been chosen to be $\lambda_0=2$. The value of $\lambda_\tau =5$, and time of observation is $\tau=1$. We find that although in SS protocol, the  $\av{\Delta s_{tot}}$ is greater than zero unlike in the case of the optimal protocol, a larger fraction of trajectories are less than zero in the former case.}
  \label{fig:entdist}
\end{figure}

This behaviour of the violation fractions is counter-intuitive: on one hand, it says that the total entropy change for the optimal protocol is minimum on average, on the other hand it says that the corresponding violation fraction can also become less.
This trend can be understood from figure \ref{fig:entdist}, where we have plotted the distributions of total entropy change for the SS protocol, where the value of $\lambda_0$ has been chosen to be equal to 2, at which $f_A$ for the SS protocol exceeds that for the optimal protocol. We clearly observe that the peak of the distribution is in the negative side for the SS protocol, which leads to a large value of $f_A$. Nevertheless, it also has a long tail that ensures that the mean of the distribution is positive. We also note that the distribution of $\Delta s_{tot}$ for the optimal protocol is actually a delta-function, as shown in Eq. \eqref{s_optimal} below.

\subsection{Analytically obtained moments of work and total entropy change}

Since the protocol is a constant with jumps and the initial and final times, the moment generating function of both $W$ and $\Delta s_{tot}$ can be calculated analytically. We first note that $W$ is simply the sum of energy changes at the terminal instants:
\begin{align}
  W = \frac{1}{2}(\lambda-\lambda_0)x_0^2 + \frac{1}{2}(\lambda_\tau-\lambda)x_\tau^2,
  \label{W_optimal}
\end{align}
where $\lambda = 1/\sigma_0^2$ (see Eq. \eqref{optimal_protocol}).
The corresponding moment generating function is given by
\begin{align}
  M_W(\alpha) = \av{e^{\alpha W}} = \int dx_0 dx_\tau p_0(x_0)P(x_\tau|x_0)e^{\alpha W(x_0,x_\tau)}.
  \label{Mw}
\end{align}
This is a Gaussian integral, with
\begin{align}
  p_0(x_0) &= \frac{1}{\sqrt{2\pi\sigma_0^2}}e^{-x_0^2/2\sigma_0^2};\nn\\
  P(x_\tau|x_0) &= \sqrt{\frac{\lambda}{2\pi (1-e^{-2\lambda \tau})}}\cdot \exp\left[-\frac{\lambda}{2}\frac{(x_\tau - x_0 e^{-\lambda \tau})^2}{1-e^{-2\lambda\tau}}\right].
                  \label{probs}
\end{align}
Plugging \eqref{probs} into \eqref{Mw} and performing the integral leads to
\begin{align}
  M_W(\alpha) &= \frac{\lambda}{\sqrt{\lambda B(\alpha)-\alpha (\lambda_\tau-\lambda)(1-e^{-2\lambda\tau})B(\alpha)-\frac{\lambda^2 e^{-2\lambda\tau}}{(1-e^{-2\lambda \tau})}}},
\end{align}
where
\begin{align}
  B(\alpha) = \lambda-\alpha(\lambda-\lambda_0)+\frac{\lambda e^{-2\lambda\tau}}{1-e^{-2\lambda\tau}}.
\end{align}
The $n^{th}$ moment of $W$ is obtained by using the relation
\begin{align}
  \av{W^n} = \frac{\partial^n}{\partial\alpha^n}M_W(\alpha)\bigg|_{\alpha=0}.
\end{align}
Using this, for $\tau=1$, $\lambda_0 = 2$ and $\lambda_\tau = 5$, we obtain the first and second moments of $W$ to be 0.75 and 1.69 respectively, which agree with our numerical results to a good accuracy.

On the other hand, the expression for total entropy change is given by
\begin{align}
  \Delta_is &= W-\Delta F - \Delta i\nn\\
            &= \frac{1}{2}\bigg(\lambda-\frac{1}{\sigma_0^2}\bigg)x_0^2 + \ln\bigg(\frac{\sigma_\tau}{\sigma_0}\bigg)+\frac{1}{2}\bigg(\frac{1}{\sigma_\tau^2}-\lambda\bigg)x_\tau^2\nn\\
            &= 0,
              \label{s_optimal}
\end{align}
keeping in mind that $\lambda = 1/\sigma_0^2$ and $\sigma_\tau=\sigma_0$, since the form of the distribution does not change during the protocol. Thus we find that even though the process is not quasistatic, we still obtain vanishing total entropy for \emph{each} individual trajectory for the optimal protocol.

\section{Fluctuation theorem for work in presence of nonequilibrium initial distribution}

We have, from \eqref{trajectory_equation},
\begin{align}
  \Delta s_{tot} = \frac{W-\Delta F}{T}-\Delta i.
\end{align}
 As has been shown in \cite{sei05_prl,sei08_epjb}, this total entropy change follows the fluctuation theorem
\begin{align}
  \av{e^{-\Delta s_{tot}}} = 1.
\end{align}
 Therefore, we obtain the following fluctuation theorem in terms of work done:
\begin{align}
  \av{e^{-\beta(W-\Delta F)+\Delta i}}=1,
\end{align}
$\beta$ being the inverse temperature of the bath.

\section{Conclusions}

In this paper, we have studied the generalization of the Maximum Work Theorem when the initial distribution is non-equilibrium. We have derived formal expressions for the violation fraction corresponding to the new inequality when the confining potential is harmonic and its centre is dragged according to a time-dependent protocol. In particular, if the centre is dragged with uniform velocty, then these formal expressions can be analytically calculated. Further, we have derived functional forms of the optimal protocol in two cases: (i) when the centre of the harmonic trap is dragged, and (ii) when the stiffness constant of the trap is changed with time. In the first case, we found that the optimal protocol is linear in time with jumps at the initial and the final points. In the second case, we found that the protocol simply consists of jumps at the end points without any time dependence in-between. We further noted that a similar protocol will be the optimal one even for more general forms of the potential (other than the harmonic case that we have treated here).
These optimal protocols are different from the ones provided in \cite{sei07_prl} even when the initial distribution is thermal, because of the nonequilibrium final distribution. Surprisingly, it was observed that although the average of the total entropy change in the optimal protocol provided above is zero, which is less than the average entropy change computed using the optimal protocol of \cite{sei07_prl}, the  latter can generate a higher fraction of trajectories that are atypical with respect to the MET in some parameter range. Which process is more efficient: one with smaller value of $\av{\Delta s_{tot}}$ or one with higher violation fraction, is unclear.
\vspace{0.5cm}

\section{Acknowledgements}

One of us (AMJ) thanks DST, India for financial support (through J. C. Bose National Fellowship).

%

\end{document}